\documentclass[10pt,twocolumn,letterpaper]{article}

\usepackage{wacv}
\usepackage{times}
\usepackage{epsfig}
\usepackage{graphicx}
\usepackage{amsmath}
\usepackage{amssymb}
\usepackage{xcolor}
\usepackage{textcomp}
\usepackage[pagebackref=true,breaklinks=true,colorlinks,bookmarks=false]{hyperref}
\usepackage{calc}
\usepackage{tabularx}
\usepackage{booktabs}
\usepackage{ragged2e}
\usepackage{xcolor}
\usepackage{multirow}
\usepackage{commath}
\usepackage{wrapfig,lipsum,booktabs}
\usepackage[accsupp]{axessibility}

\definecolor{blue1}{RGB}{0,0, 122}

\def\wacvPaperID{461} % Enter the WACV Paper ID here

\wacvfinalcopy % *** Uncomment this line for the final submission

\ifwacvfinal
\def\assignedStartPage{1} % *** Enter the assigned starting page number (instead of 9876)
\fi

\ifwacvfinal
\usepackage[breaklinks=true,bookmarks=false]{hyperref}
\else
\usepackage[pagebackref=true,breaklinks=true,colorlinks,bookmarks=false]{hyperref}
\fi

\ifwacvfinal
\else
\fi

\begin{document}

%%%%%%%%% TITLE
\title{No-Reference Image Quality Assessment via Transformers, Relative Ranking, and Self-Consistency}

\author{S. Alireza Golestaneh\textsuperscript{\rm 1}\thanks{Currently at Bosch Center for AI.} \quad  Saba Dadsetan \textsuperscript{\rm 2}  \quad Kris M. Kitani\textsuperscript{\rm 1}\\
\textsuperscript{\rm 1} Carnegie Mellon University \quad  \textsuperscript{\rm 2} University of Pittsburgh}

\maketitle

%%%%%%%%% ABSTRACT
\begin{abstract}
The goal of No-Reference Image Quality Assessment (NR-IQA) is to estimate the perceptual image quality in accordance with subjective evaluations, it is a complex and unsolved problem due to the absence of the pristine reference image. 
In this paper, we propose a novel model to address the NR-IQA task by leveraging a hybrid approach that benefits from Convolutional Neural Networks (CNNs) and self-attention mechanism in Transformers to extract both local and non-local features from the input image.
We capture local structure information of the image via CNNs, then to circumvent the locality bias among the extracted CNNs features and  obtain a non-local representation of the image, we utilize Transformers on the extracted features where we model them as a sequential input to the Transformer model. 
Furthermore, to improve the monotonicity correlation between the subjective and objective scores,  we utilize the relative distance information among the images within each batch and enforce the relative ranking among them. 
Last but not least, we observe that the performance of NR-IQA models degrades when we  apply equivariant transformations (\textit{e.g.} horizontal flipping) to the inputs.
Therefore, we propose a method that leverages self-consistency as a source of self-supervision to improve the robustness of NR-IQA models. 
Specifically, we enforce self-consistency between the outputs of our quality assessment model for each image and its transformation (horizontally flipped) to utilize the rich self-supervisory information and reduce the uncertainty of the model. 
To demonstrate the effectiveness of our work, we evaluate it on seven standard IQA datasets (both synthetic and authentic) and show that our model achieves state-of-the-art results on various datasets. \footnote{Code will be released  \href{https://github.com/isalirezag/TReS}{here.}}
\end{abstract}

\vspace{-0.5 cm}
\section{Introduction}
Being able to predict the perceptual image quality robustly and accurately  without having access to the   reference image is crucial for different computer vision applications as well as social and streaming media industries.
On a routine day, on average  several   billion photos are uploaded and shared on social media platforms such as Facebook, Instagram, Google, Flicker, \textit{etc}; low-quality images can serve as irritants when they convey a negative impression to the viewing audiences.
On the other hand, at one extreme, not being able to assess the image  quality accurately can be life-threatening (\textit{e.g.,} when low-quality images impede the ability of autonomous vehicles \cite{lou2019veri,chiu2020assessing} and traffic controllers \cite{zhu2016traffic} to safely navigate environments).

Objective image quality assessment (IQA) attempts to use computational models to predict the image quality  in a manner that is consistent with quality ratings provided by human subjects.
Objective quality metrics can be divided into full-reference (reference available or FR), reduced-reference (RR), and no-reference (reference not available or NR) methods based on the availability of a reference image \cite{wang2006modern}.
The goal of the no-reference image quality assessment (NR-IQA) or blind image quality assessment (BIQA) methods is to provide a solution when the reference image is not available \cite{saad2012blind,mittal2012no,xue2013learning,zhang2015feature}.

NR-IQA  mainly divides into two groups, distortion-based and general-purpose methods. 
A distortion-based approach aims to predict the quality score for a specific type of distortion (\textit{e.g.,} blocking, blurring).
Distortion-based approaches have limited applications in \textit{real-world} scenarios since we cannot always specify distortion types.
Thus, a general-purpose approach is designed to evaluate image quality without being limited to distortion types.
General-Purpose methods make use of extracted features that are informative for various types of distortions.
Therefore, their performances highly depend on designing elaborate features.

Traditionally, general-based NR-IQA methods focused on quality assessment for synthetically distorted images (\textit{e.g.,} Blur, JPEG, Gaussian Noise). 
However, the main challenges along with existing synthetically distorted datasets are 1) they contain limited content and distortion diversity, and 2) they do not capture complex mixtures of distortions that often occur in \textit{real-world} images. 
Recently, by introducing more \textit{in-the-wild} datasets such as
CLIVE \cite{ghadiyaram2015massive},  KonIQ-10K \cite{hosu2020koniq}, and LIVEFB \cite{ying2019patches} we can have a better understanding of complex  distortions (\textit{e.g.,} poor lighting conditions,
sensor limitations, lens imperfections, amateur manipulations) that often occur in
\textit{real-world} images. 
In contrast to  synthetic distortions in which   
degradation processes are precisely specified and can be simulated in laboratory environments, authentic  distortions are more complicated because there is no reference-image  
available, and it is unclear how the human visual system  (HVS) distinguishes between the picture quality and picture authenticity.
For instance, while distortion can detract from
aesthetics, it can also contribute to it, as when intentionally
adding blur (bokeh) to achieve photographic effects.
Moreover, HVS perceives image quality differently among various image contents and quantifies image quality differently  for images with different contents with the same level and type  of distortion  \cite{chou1995perceptually,wang2009mean,alam2014local,li2018has}.

Existing deep learning-based IQA methods mainly rely only on the subjective human scores (MOS/DMOS) and modeling the quality prediction task mainly as a regression or classification task.
This causes the models not   to be able to  leverage the relative ranking between the images explicitly.
We propose to take into account the relative distance information between the images within each batch and enforce our model to learn the relative ranking between the images with the highest and lowest quality scores in addition to the quality assessment task.

Moreover, as shown in Fig. \ref{F1}, 
 despite using common augmentation techniques during the training, the performance of IQA methods degrade
when we apply a simple equivariant transformation  (\textit{e.g.,} horizontal filliping) to the input image.
This contradicts the way that humans perceive the quality of images. 
In other words, subjective perceptual quality scores remain the same for specific equivariant transformations that can appear very often during real-life applications.
To alleviate this issue we propose a self-consistency approach that enforces our model to have consistent predictions for an image and its transformed version.
The contributions of this work are summarized as follows:

\begin{figure}[t]
\centering
  \includegraphics [scale=.3 ]{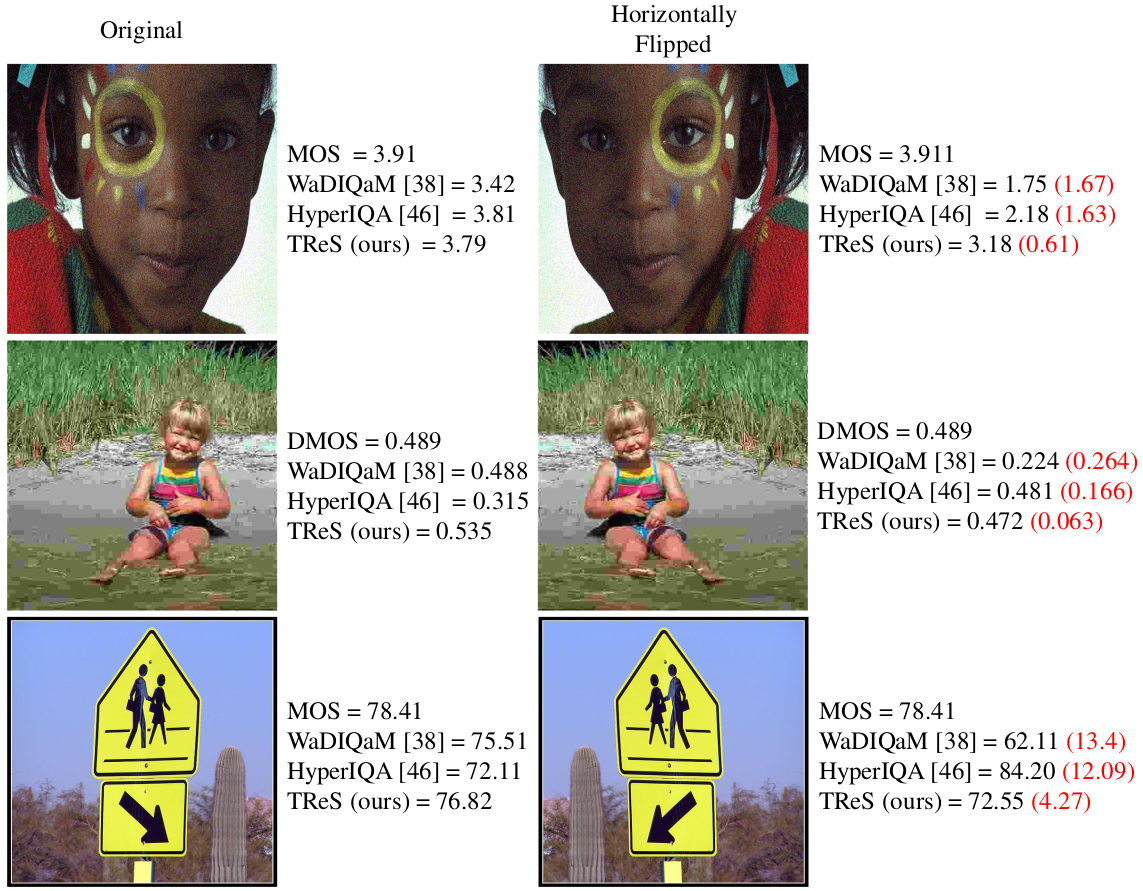}
	\caption{Illustration of the sensitivity of NR-IQA models to horizontal flipping. 
	On the right side of each image, we provide the subjective quality score (MOS/DMOS) and the predicted quality score; the red numbers in the parentheses show the absolute difference between the predictions when the image is flipped. 
}
	\label{F1}
 \end{figure}

\begin{itemize}
\item We introduce an end-to-end deep learning approach for NR-IQA. Our proposed model utilizes local and non-local information of an image by leveraging CNNs and self-attention mechanism of Transformers.
Particularly, in addition to local features that are generated via CNNs, we take advantage of the sequence modeling and self-attention mechanism of   Transformers to learn a non-local representation of the image from the multi-scale features that are extracted from different layers of CNNs.
The non-local features are then fused with the local features to predict the final image quality score   (Sec. \ref{S3.1},  \ref{S3.2}, and \ref{S3.3}).

\item We propose a relative ranking loss that explicitly enforces the relative ranking among the samples.
We propose to use a triplet loss with an adaptive margin based on the human subjective scores to make the distance between the image with the highest (lowest) quality score closer to the one with the second-highest (second-lowest) quality score and further away from the image with the lowest (highest)  score  (Sec. \ref{S3.4}).

\item Lastly, we propose to use an equivariant transformation of the input image as a source of self-supervisory to improve the robustness of our proposed model.
During the training,  we use  self-consistency
between the output for each image and its transformation to utilize the rich self-supervisory information and reduce the sensitivity of the network  (Sec. \ref{S3.5}).

\item Extensive  experiments on seven benchmark datasets (for both authentic and synthetic distortions) confirm that our proposed method
performs well across different datasets.

\end{itemize}

\section{Related Work}
Before the raise of deep learning, early version of general-purpose NR-IQA methods  mainly   divide  into natural scene statistics (NSS) based metrics 
\cite{moorthy2010two,moorthy2011blind,mittal2012no,saad2012blind,mittal2012making,gao2013universal,
zhang2015feature,xu2016blind,ghadiyaram2017perceptual} and learning-based metrics 
\cite{ye2012no,ye2012unsupervised,zhang2014training,ye2014beyond,zhang2015som,ma2017dipiq}. 
The underlying assumption  for hand-crafted feature-based approaches is that the  natural scene statistics (NSS) extracted from natural images are highly regular \cite{simoncelli2001natural} and different distortions will break such statistical regularities. 
 Variations of NSS features in different domains such as  spatial   \cite{mittal2012no,mittal2012making,zhang2015feature}, gradient  \cite{zhang2015feature}, discrete cosine transform (DCT)    \cite{saad2012blind}, and wavelet  \cite{moorthy2011blind}, showed impressive performances for synthetically distorted images.
Learning-based approaches  utilize  machine learning techniques such as dictionary learning  to map the learned features to the human subjective scores.
Ye \textit{et al.} \cite{ye2012unsupervised} used a dictionary learning method to encode the raw image patches to  features and to predict subjective  quality scores by support vector regression (SVR) model.
 Zhang \textit{et al.} \cite{zhang2015som} combined the semantic-level features  with local features for   quality estimation. 
Although early versions of hand-crafted and feature learning methods perform well on small synthetically distorted datasets, they suffer from not being able  to model \textit{real-world} distortions.

\textbf{Deep learning for NR-IQA.} By  success of deep learning \cite{krizhevsky2012imagenet,he2016deep} in many computer vision tasks, different approaches  utilize deep learning for  NR-IQA
\cite{kang2014convolutional,long2015fully,redmon2016you,bosse2017deep,zeng2017probabilistic,lin2018hallucinated,talebi2018nima,ma2017end,zhang2018blind,lin2018hallucinated,bianco2018use,yan2019naturalness,su2020blindly,zhu2020metaiqa,hosu2020koniq,zhang2020learning}.
Early version of deep learning NR-IQA methods  \cite{kang2014convolutional,bosse2016deep,kim2017deep,bosse2017deep,kim2017deep,zeng2017probabilistic} leveraged   deep features from  CNNs \cite{krizhevsky2012imagenet,he2016deep} while  pretrained on large classification dataset ImageNet \cite{deng2009imagenet}.
\cite{kang2015simultaneous,xu2016multi}   addressed NR-IQA  in a multi-task manner where they leverage subjective quality score as well as distortion type simultaneously during the training.
 Ma \textit{et al.} \cite{ma2017end} proposed  a multi-task network  where  two sub-networks train in two stages for distortion identification  and  quality prediction.
\cite{kim2016fully,kim2018deep,pan2018blind,lin2018hallucinated} used some sort of the reference images during their training  to predicted the quality score in a blind manner.
Hallucinated-IQA \cite{lin2018hallucinated} proposed an  NR-IQA method based on generative adversarial models, where they first generated a hallucinated reference image to compensate for 
the absence of the true reference and then paired the information of hallucinated reference with the distorted image to estimate the quality score.
Talebi \textit{et al.} \cite{talebi2018nima} proposed a CNN-based model to predict the perceptual distribution of subjective quality  scores (instead of the mean value). 
Zhu \textit{et al.} \cite{zhu2020metaiqa} proposed a model to leverage meta-learning to   learn the prior knowledge that is shared among different distortion types.
Su \textit{et al.} \cite{su2020blindly} proposed a model that extracts content features from the deep model in different scales and pools them to predict image quality.

\textbf{Transformers for NR-IQA.} 
Currently, CNNs are the main backbone for  features extraction among the state-of-the-art NR-IQA models.
Although CNNs capture the local structure of the image, they are well known for missing to capture non-local information and having strong locality bias.
Furthermore, CNNs demonstrated  a bias towards spatial invariance through  shared weights across all positions which makes them  ineffective if a more complex combination of features is needed.
Since IQA   highly depends on both local and non-local features, we propose to use Transformers and CNNs together.
Inspired by NLP which widely employs Transformers block to   model long-range dependencies in language sequence, we utilize a Transformer-based network to compute the dependencies among the CNN extracted features from multi-scales and model the non-local dependency among the extracted features.

Transformers were introduced by Vaswani \textit{et al.} \cite{vaswani2017attention} as a new attention-based
building block for machine translation. 
Attention mechanisms \cite{bahdanau2014neural}  are neural network layers that aggregate information from the entire input sequence. 
Due to the success of Transformer-based models in the
NLP field, we start to see different attempts to explore the benefits
of Transformer for computer vision tasks \cite{chen2020rewon,dosovitskiy2020image,dosovitskiy2020image,chen2020pre,han2020survey,carion2020end}.
The application of Transformer for NR-IQA is not explored yet.
Concurrently with our work, \cite{you2020transformer} used Transformers for NR-IQA, where the features from the last layer of CNNs were sent to  Transformers for the quality prediction task.
Different from
 \cite{you2020transformer} that use Transformers as an additional  feature extraction block at the end of CNNs, we use it to model the non-local dependency between the extracted  multi-scale features.
Notably, we leverage the temporal sequence modeling properties of Transformers to compute a non-local representation of the image from the multi-scale features (Sec. \ref{S3.2}).

\begin{figure*}[!t]
\centering
  \includegraphics [scale=.32 ]{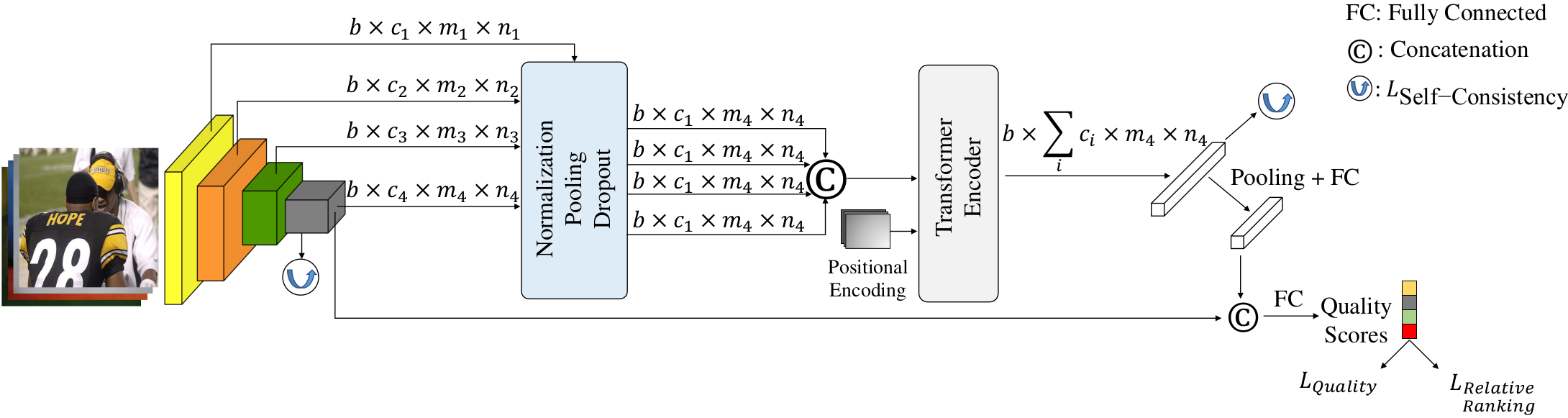}
	\caption{Flowchart of our proposed NR-IQA algorithm.}
	\label{F2}
 \end{figure*}

\textbf{Learning to rank for NR-IQA.} 
These approaches \cite{gao2015learning,liu2017rankiqa,ma2017end,zhang2018blind} address NR-IQA as a learning-to-rank problem, where  the relative ranking information is used during the training.
 Zhang \textit{et al.} \cite{zhang2018blind} leveraged discrete ranking information from images of the same content and distortion but at different levels (degree of distortion) for quality prediction.  
 \cite{zhang2021uncertainty}   used continuous ranking information from MOSs and variances between the subjective scores.
\cite{ma2017end,ma2019blind} extracted binary ranking information via FR-IQA methods during the training. However, due to the use of reference images, their method is only applicable to    synthetic distortions.
Existing ranking-based algorithms use a fixed margin (which is selected empirically) to minimize their losses.
Also, most of the aforementioned approaches (except  \cite{zhang2021uncertainty}) fails to perform well on the authentic datasets mainly due to the requirement of using referee images during the training stage.
In our proposed method, we also leverage the MOS/DMOS  information for relative ranking. However, in contrast to the existing methods, we propose to minimize the relative distance among the samples via a triplet loss with an adaptive margin which does not need the empirical margin selection.

\textbf{Poor generalization in deep neural
networks} is a well-known problem and an active area of research. 
In IQA tasks poor generalization is mostly considered as when  the model performs well on the  dataset that it is trained on but poorly on another dataset with the same type of artifacts.
Reasons such as various contents, domain shift, or scale shift in the subjective scores mainly cause the poor generalization of IQA models.
In our experiments, in addition to cross dataset evaluation, we also notice that the performance of  deep-learning based IQA models degrades when we  apply horizontal flipping or rotation to the inputs.
Common approaches such
as dropout \cite{srivastava2014dropout}, ensembling \cite{zhou2012ensemble}, cross-task consistency \cite{zamir2020robust}, and augmentation proposed to increase the generalization
in deep models. 
However,  making predictions using a whole
ensemble of models is cumbersome and   too computationally expensive  due to large models' sizes.
Data augmentation has been successful in increasing the
generalization of CNNs. 
However, as shown in Fig. \ref{F1}, a model can still suffer from poor
generalization for a simple transformation. 
\cite{worrall2017harmonic, cohen2016group} show that although using different augmentation methods improve the generalization of CNNs, they are still
sensitive to equivariant perturbations in data.

As shown in Fig. \ref{F1},   NR-IQA models that used
image flipping as an augmentation during training still fail to have a robust quality prediction for
an image and its flipped version. This kind of high variance in the quality prediction can affect the robustness of computer vision applications directly.
In this work, we improve the consistency of our model via the simple observation that the results of the NR-IQA model should not change under transformations such as horizontal flipping.

\vspace{-0.22 cm}
\section{Proposed Method}
In this section, we detail our proposed model, which is an NR-IQA method based on \textbf{T}ransformers, \textbf{Re}lative ranking, and \textbf{S}elf consistency, namely \textit{\textbf{TReS}}.
Fig. \ref{F2} shows an overview of our proposed method.

\subsection{Feature Extraction}\label{S3.1}
Given an input image $I\in \mathbb{R}^{3\times m\times n}$, where $m$ and $n$ denote width and height, our goal is to estimate its perceptual quality score ($q$). 
Let $f_{\phi}$ represent a CNN  with learnable parameters $\phi$, and  $F_{i}\in\mathbb{R}^{b\times c_{i}\times m_{i}\times n_{i}}$ denotes the features from the $i^{th}$ block of CNN, where  $i\in\{1,2,3,4\}$,  $b$ denotes the batch size, and $c_{i}, m_{i}$, and $ n_{i}$ denote the  channel size, width, and height of the $i^{th}$ feature, respectively.
Let $F_{4}\in\mathbb{R}^{b\times c_{4}\times m_{4}\times n_{4}}$ represent  the high-level semantic features from the last layer in the $4^{th}$ block of CNN.
We use the last layers of each block to extract the multi-scale features from the input image.
Since the extracted features from different layers have different ranges, statistics, and dimensions,   we first send them to normalization, pooling, and dropout layers.
For normalization and pooling we use Euclidean norm which is defined by $F_{i}=\frac{F_{i}}{\max(\left\Vert F_{i}\right\Vert _{2},\epsilon)}$ followed by a $\mathit{l}_2$ pooling layer   \cite{henaff2015geodesics,ding2020image} which has been used to demonstrate the behavior of complex cells in primary 
visual cortex \cite{bruna2013invariant,vintch2015convolutional}.
The $\mathit{l}_2$ pooling layer defines by:
\begin{equation}
P(x)=\sqrt{g\ast(x\odot x)},
\end{equation}
where $\odot$ denotes point-wise product, and the blurring kernel
$g(.)$ is implemented via a Hamming window that approximately applies the Nyquist criterion \cite{ding2020image}.
Let $\bar{F}_{i}\in\mathbb{R}^{b\times c_{i}\times m_{4}\times n_{4}}$ denote the output feature after sending $F_i$  to the normalization, pooling, and dropout layers.
Next, we  concatenate $\bar{F}_{i}$, where $i\in \{1,2,3,4\}$, and  denote  the output by $
\tilde{F}\in\mathbb{R}^{b\times\sum_{i}c_{i}\times m_{4}\times n_{4}}$.

\subsection{Attention-Based Feature Computation}
\label{S3.2}
CNNs exploit  the structure of images via local interactions through convolution with small kernel sizes. 
Different layers of a network can have different semantic information that is   captured through the local interactions,  and as we move from lower layers to the higher layers, the computed features carry more semantic about the content of the image \cite{zeiler2014visualizing}.
IQA depends on both low- and high-level features. 
A model that does not take into account both low- and high-level features  can  mistake a plain sky as a low-quality image \cite{li2018has}.
Moreover, due to the architecture of CNNs   they mainly capture the local spatial structure of the image and are unable to  model the relation  among  the  non-local features. 

\begin{figure}[t]
\centering
  \includegraphics [scale=.55 ]{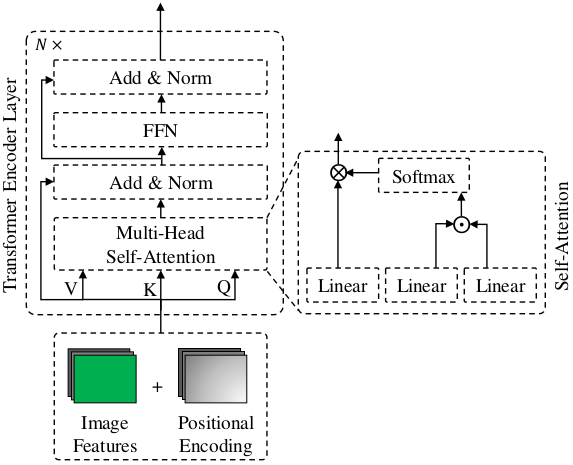}
	\caption{ Illustration of multi-head multi-layer self-attention module of the Transformer  Encoder layer. $N$ is a hyperparameter that denotes the number of encoder layers In the Transformer which stack together.}
	\label{F3}
 \end{figure}

Transformers have shown impressive results in modeling the dependencies among the sequential data.
Therefore, we use the encoder part of the Transformer, which is composed of a multi-head self-attention layer and a feed-forward neural network \cite{sutskever2014sequence,vaswani2017attention,carion2020end} to perform attention operations over the multi-scale extracted features from different layers   and model the dependencies among them.
We follow the encoder architecture of \cite{carion2020end} (see Fig. \ref{F3}).
We model the features from different layers of CNN as a sequence of information ($\tilde{F}$) and send them to the Transformer encoder.
Since the self-attention mechanism is a non-local operation, we use it to compute a non-local representation of the image.
In other words, we use Transformers to compute  information for each element of extracted features with respect to the others (not only the local neighbor features).  
Transformer architecture contains no built-in inductive prior to the locality of interactions and,  is free to learn complex relationships across the features. 
We also add positional encoding (in a similar way as \cite{bello2019attention,parmar2018image}) to the input of the attention layers to deal with the permutation-invariant property of Transformers.
The positional encoding will also let our model be aware of the position of the features that contribute the most to the IQA task.

In detail, given an input ($\tilde{F}$) and the number of heads ($\mathit{h}$), the input  is first transformed into three different
groups of vectors,  the query group, the key group and
the value group. 
Given a multi-head attention module with $\mathit{h}$ heads and  dimension of $d$, each of the aforementioned groups will have dimension of $d^{'}=\frac{d}{h}$.
Then, features derived from different inputs are packed together
into three different groups of matrices $Q',K',$ and $V'$, where $Q' = \{Q_{i}\}_{i=1}^{h}=Concat(Q_{1},...,Q_{h})$ and the same definition applies to $K'$ and $V'$. 
Next, the process of multi-head attention is computed as follows:
\vspace{-0.2 cm}
\begin{equation}
\resizebox{.9\hsize}{!}{${MultiHead}(Q',K',V')=Concat({head}_{1},...,{head}_{h})W_1$}
\end{equation}
where $\mathit{W_1}$ is   the linear projection matrix and has dimension of $d\times d$, ${head}_{i}={Attention}(Q_{i},K_{i},V_{i}),$ and  $Attention(Q,K,V)=softmax(\frac{Q K^{T}}{\sqrt{d'}})\odot V$.
For the first Transformer encoder layer, the query, key, and value matrices ($Q$, $K$, and $V$) are all the same.
Following the encoder architecture design in \cite{vaswani2017attention}, to strengthen the flow of information and improve the performance,  a residual connection followed by a layer normalization is added in each sub-layer in the encoder.
Next, a feed-forward network (FNN) is
applied after the self-attention layers \cite{vaswani2017attention}. 
FNN is consist of two
linear transformation layers and a ReLU activation function
within them, which can be denoted as  $FFN(X) = {W_3}\sigma({W_2}X + b'_2) + b'_3$,
where $\mathit{W_2}$ and $\mathit{W_3}$ are the two parameter matrices, and $b'_2$ and $b'_3$ are the biases. $\sigma$ represents the ReLU activation function.
Let $\hat{F}\in\mathbb{R}^{b\times\sum_{i}^{4}c_{i}\times m_{4}\times n_{4}}$ represent the output features from the final Transformer encoder layer.

\subsection{Feature Fusion and Quality Prediction}
\label{S3.3}
To benefit from the extracted features from both local (convolution) and non-local (self-attention) operators 
we use fully connected (FC) layers  as  fusion layers to map the aforementioned features and predict the perceptual quality of the image (see Fig. \ref{F2}).
For each batch of images, $B$, we minimize the  regression loss to train our network.
\begin{equation}
\label{E3}
\resizebox{.6\hsize}{!}{$\mathcal{L}_{Quality,B}=\frac{1}{N}\sum_{i}^{N}\left\Vert q_{i}-s_{i}\right\Vert,$}
\end{equation}
where $q_i$ is the predicted quality score for $i^{th}$ image and $s_i$ is its corresponding ground truth (subjective  quality score).
 
\subsection{Relative Ranking}
\label{S3.4}
Although the regression loss (Eq. \ref{E3}) is effective for the quality prediction task, it does not explicitly take into account ranking and correlation among images.
Here, our goal is to consider the relative ranking relation between the samples within each batch.
It is computationally expensive to consider \textit{all} the samples' relative ranking information; therefore, we only enforce it for the extreme cases.
 Among  images within the batch  $B$, let $q_{max},q'_{max},q_{min},$ and $q'_{min}$ denote the predicted quality for images with highest, second highest, lowest, and second lowest subjective quality scores, respectively, \textit{i.e.,}  $s_{q_{max}}>s_{q'_{max}}>s_{q'_{min}}>s_{q_{min}}$, where $s_{q_{max}}$ denotes the subjective quality score corresponding to the image with the predicted quality score $q_{max}$, and a similar notation rule applies to the rest.
Our goal is to have $d(q_{max},q'_{max})+margin_1\leq d(q_{max},q_{min})$, here we define, $d(x,y)$ as the absolute value between $x$ and $y$,  $d(x,y)=\abs{x-y}$. 
We utilize triplet loss to address the above inequality, where we minimize $\max\{0,d(q_{max},q'_{max})-d(q_{max},q_{min})+margin_{1}\}$.
In a similar way, we also want to have $d(q_{min},q'_{min})+margin_2 \leq d(q_{max},q_{min})$.
The $margin$ values can be selected empirically based on each dataset, but that is cumbersome since each dataset have different distributions and ranges for the quality scores.
For a perfect prediction where the estimated quality scores are the same as subjective scores we will have $margin_1+\abs{s_{q_{max}}-s_{q'_{max}}}\leq \abs{s_{q_{max}}-s_{q_{min}}} \rightarrow margin_1 +  (s_{q_{max}}-s_{q'_{max}}) \leq (s_{q_{max}}-s_{q_{min}})  \rightarrow margin_1 \leq s_{q'_{max}}  -s_{q_{min}}.$
Therefore, we can consider $s_{q'_{max}}  -s_{q_{min}}$ to be an upper-bound for $margin_1$ during the training, and set  $margin_1 = s_{q'_{max}}  -s_{q_{min}}$ in Eq. \ref{eee4}.
Similarly,  we define $ margin_2=s_{q_{max}}-s'_{q_{min}}$. 
Finally, our relative ranking loss is defined as:
\vspace{-0.2 cm}
\begin{equation}
  \begin{aligned}
     & \mathcal{L}_{Relative-Ranking,B} =\\
     & \mathcal{L}_{triplet}(q_{max},q'_{max},q_{min})+\mathcal{L}_{triplet}(q_{min},q'_{min},q_{max})\\
     & =\max\{0,d(q_{max},q'_{max})-d(q_{max},q_{min})+margin_{1}\}\\   
     & +\max\{0,d(q'_{min},q_{min})-d(q_{max},q_{min})+margin_{2}\}.
    \end{aligned}
    \label{eee4}
\end{equation}

\vspace{-0.2 cm}
\subsection{Self-Consistency}
\label{S3.5}
Last but not least, we propose to utilize the model's uncertainty for the
input image and its equivariant transformation during the training process.
We exploit  self-consistency via the self-supervisory signal between
each image and its equivariant transformation  to increase the robustness 
of the model. 
Let for an input $I$, $f_{\phi,conv}(I)$ and $f_{\theta,atten}(I)$  denote the output logits  
 belonging to outputs of the convolution   and Transformer layers, respectively, where $f_{\phi,conv}$ and $f_{\theta,atten}$ represent the CNN and Transformer  with learnable parameters $\phi$ and $\theta$, respectively. 
 In our model, we use the outputs of  $f_{\phi,conv} $ and $f_{\theta,atten}$ to predict the image quality and since the human subjective scores stay the same for the horizontal filliping version of the input image, we thus expect to have $f_{\phi,conv}(I)=f_{\phi,conv}(\tau(I))$ and $f_{\theta,atten}(I)=f_{\theta,atten}(\tau(I))$, where $\tau$ represents the horizontal filliping transformation. 
 In this way, by applying our consistency loss, the network  learns to reinforce representation learning of itself without   additional labels and external supervision.
We minimize the self-consistency loss that is defined as follows:
% \vspace{-0.1 cm}
\begin{equation}
  \begin{aligned}
     & \mathcal{L}_{Self-Consistency}=\\
     & \left\Vert f_{\phi,conv}(I)-f_{\phi,conv}(\tau(I))\right\Vert +\\
     & \left\Vert f_{\theta,atten}(I)-f_{\theta,atten}(\tau(I))\right\Vert +\\
     & \lambda_{1}\left\Vert \mathcal{L}_{Relative-Ranking,B}-\mathcal{L}_{Relative-Ranking,\tau(B)}\right\Vert,
    \end{aligned}
\end{equation}
where $\tau(B)$ denote when   the equivariant transformation applies on  batch $B$.

\subsection{Losses}
Our model trains in an end-to-end manner and minimizes the aforementioned losses together simultaneously.
The total  loss for our   model is defined as:
\begin{equation}
\begin{array}{lll}
\mathcal{L}_{total}& =\mathcal{L}_{Quality}+\lambda_{2}\mathcal{L}_{Relative-Ranking}+\\
& \lambda_{3} \mathcal{L}_{Self-Consistency},
\end{array}
\end{equation}
where $\lambda_{1},\lambda_{2},\lambda_{3}$  are balancing coefficients.

\section{Experiments}
\subsection{Datasets and  Evaluation Metrics}
We evaluate the performance of our proposed model extensively on seven publicly available IQA datasets (four synthetically distorted and three authentically distorted).
For synthetically distorted datasets, we use LIVE \cite{sheikh2006statistical}, CSIQ \cite{larson2010most}, TID2013 \cite{ponomarenko2015image}, and KADID-10K \cite{lin2019kadid}, where among them  KADID  has the most number of distortion typesand  distorted images.
For authentically distorted datasets, we use CLIVE \cite{ghadiyaram2015massive},  KonIQ-10k \cite{hosu2020koniq}, and LIVE-FB \cite{ying2019patches}, where among them LIVE-FB has the most number of unique contents.
Table \ref{TB0} shows the summary of the datasets that are used in our experiments.
% \vspace{-5pt}

\begin{table}[h!]
\centering
\caption{Summary of IQA datasets.}
\resizebox{2.1 in}{!} {
\begin{tabular}{cccc}
\hline 
\multirow{2}{*}{Databases} & \# of Dist. & \# of Dist.  & Distortions \tabularnewline
 & Images & Types & Type\tabularnewline
\hline 
LIVE & 799 & 5 & synthetic\tabularnewline
CSIQ & 866 & 6 & synthetic\tabularnewline
TID2013 & 3,000 & 24 & synthetic\tabularnewline
KADID & 10,125 & 25 & synthetic\tabularnewline
CLIVE & 1,162 & - & authentic\tabularnewline
KonIQ & 10,073 & - & authentic\tabularnewline
LIVEFB & 39,810 & - & authentic\tabularnewline
\hline 
\end{tabular}
}
\label{TB0}

\end{table}

\begin{table*}[t!]
\centering
\caption{Comparison of \textit{TReS} v.s. state-of-the-art NR-IQA algorithms on synthetically and authentically distorted datasets.
Bold entries in \textbf{black} and  \textbf{\textcolor{blue1}{blue}} are the best and second-best performers, respectively. * code were not available publicly.}
\resizebox{6.8 in}{!} {

\begin{tabular}{c||cc|cc|cc|cc||cc|cc|cc||cc}
\multicolumn{1}{c}{} &  & \multicolumn{1}{c}{} &  & \multicolumn{1}{c}{} &  & \multicolumn{1}{c}{} &  & \multicolumn{1}{c}{} &  & \multicolumn{1}{c}{} &  & \multicolumn{1}{c}{} &  & \multicolumn{1}{c}{} &  & \tabularnewline
\hline 
\hline 
 & \multicolumn{2}{c|}{LIVE} & \multicolumn{2}{c|}{CSIQ} & \multicolumn{2}{c|}{TID2013} & \multicolumn{2}{c||}{KADID} & \multicolumn{2}{c|}{CLIVE} & \multicolumn{2}{c|}{KonIQ} & \multicolumn{2}{c||}{LIVEFB} & \multicolumn{2}{c}{Weighted Average}\tabularnewline
\cline{2-17} \cline{3-17} \cline{4-17} \cline{5-17} \cline{6-17} \cline{7-17} \cline{8-17} \cline{9-17} \cline{10-17} \cline{11-17} \cline{12-17} \cline{13-17} \cline{14-17} \cline{15-17} \cline{16-17} \cline{17-17} 
 & PLCC & SROCC & PLCC & SROCC & PLCC & SROCC & PLCC & SROCC & PLCC & SROCC & PLCC & SROCC & PLCC & SROCC & PLCC & SROCC\tabularnewline
\hline 
HFD{*}\cite{wu2017hierarchical} & \textbf{0.971} & 0.951 & 0.890 & 0.842 & 0.681 & 0.764 & - & - & - & - & - & - & - & - & - & -\tabularnewline
PQR{*}\cite{zeng2017probabilistic} & \textbf{0.971} & 0.965 & 0.901 & 0.873 & 0.864 & 0.849 & - & - & 0.836 & 0.808 & - & - & - & - & - & -\tabularnewline
DIIVINE\cite{saad2012blind} & 0.908 & 0.892 & 0.776 & 0.804 & 0.567 & 0.643 & 0.435 & 0.413 & 0.591 & 0.588 & 0.558 & 0.546 & 0.187 & 0.092 & 0.323 & 0.264\tabularnewline
BRISQUE\cite{mittal2012no} & 0.944 & 0.929 & 0.748 & 0.812 & 0.571 & 0.626 & 0.567 & 0.528 & 0.629 & 0.629 & 0.685 & 0.681 & 0.341 & 0.303 & 0.457 & 0.430\tabularnewline
ILNIQE\cite{zhang2015feature} & 0.906 & 0.902 & 0.865 & 0.822 & 0.648 & 0.521 & 0.558 & 0.534 & 0.508 & 0.508 & 0.537 & 0.523 & 0.332 & 0.294 & 0.430 & 0.394\tabularnewline
BIECON\cite{kim2017fully} & 0.961 & 0.958 & 0.823 & 0.815 & 0.762 & 0.717 & 0.648 & 0.623 & 0.613 & 0.613 & 0.654 & 0.651 & 0.428 & 0.407 & 0.527 & 0.507\tabularnewline
MEON\cite{ma2017end} & 0.955 & 0.951 & 0.864 & 0.852 & 0.824 & 0.808 & 0.691 & 0.604 & 0.710 & 0.697 & 0.628 & 0.611 & 0.394 & 0.365 & 0.514 & 0.479\tabularnewline
WaDIQaM\cite{bosse2017deep} & 0.955 & 0.960 & 0.844 & 0.852 & 0.855 & 0.835 & 0.752 & 0.739 & 0.671 & 0.682 & 0.807 & 0.804 & 0.467 & 0.455 & 0.595 & 0.584\tabularnewline
DBCNN\cite{zhang2018blind} & \textbf{0.971} & \textbf{\textcolor{blue1}{0.968}} & \textbf{0.959} & \textbf{0.946} & 0.865 & 0.816 & \textbf{\textcolor{blue1}{0.856}} & 0.851 & 0.869 & \textbf{0.869} & 0.884 & 0.875 & 0.551 & \textbf{\textcolor{blue1}{0.545}} & 0.679 & 0.671\tabularnewline
TIQA\cite{you2020transformer} & 0.965 & 0.949 & 0.838 & 0.825 & 0.858 & 0.846 & 0.855 & 0.850 & 0.861 & 0.845 & 0.903 & 0.892 & 0.581 & 0.541 & 0.698 & 0.670 \tabularnewline
MetaIQA\cite{zhu2020metaiqa} & 0.959 & 0.960 & 0.908 & 0.899 & \textbf{\textcolor{blue1}{0.868}} & 0.856 & 0.775 & 0.762 & 0.802 & 0.835 & 0.856 & 0.887 & 0.507 & 0.540 & 0.634 & 0.656\tabularnewline
P2P-BM\cite{ying2019patches} & 0.958 & 0.959 & 0.902 & 0.899 & 0.856 & \textbf{\textcolor{blue1}{0.862}} & 0.849 & 0.840 & 0.842 & 0.844 & 0.885 & 0.872 & 0.598 & 0.526 & 0.705 & 0.658\tabularnewline
HyperIQA\cite{su2020blindly} & 0.966 & 0.962 & \textbf{\textcolor{blue1}{0.942}} & \textbf{\textcolor{blue1}{0.923}} & 0.858  & 0.840 & 0.845 & \textbf{\textcolor{blue1}{0.852}} & \textbf{0.882} & \textbf{\textcolor{blue1}{0.859}} & \textbf{\textcolor{blue1}{0.917}} & \textbf{\textcolor{blue1}{0.906}} & \textbf{\textcolor{blue1}{0.602}} & 0.544 & \textbf{\textcolor{blue1}{0.715}} & \textbf{\textcolor{blue1}{0.676}}\tabularnewline
TReS (proposed) & \textbf{\textcolor{blue1}{0.968}} & \textbf{0.969} & \textbf{\textcolor{blue1}{0.942}} & 0.922 & \textbf{0.883} & \textbf{0.863} & \textbf{0.858} & \textbf{0.859} & \textbf{\textcolor{blue1}{0.877}} & 0.846 & \textbf{0.928} & \textbf{0.915} & \textbf{0.625} & \textbf{0.554} & \textbf{0.732} & \textbf{0.685}\tabularnewline
\hline 
\hline 
\multicolumn{1}{c}{} &  & \multicolumn{1}{c}{} &  & \multicolumn{1}{c}{} &  & \multicolumn{1}{c}{} &  & \multicolumn{1}{c}{} &  & \multicolumn{1}{c}{} &  & \multicolumn{1}{c}{} &  & \multicolumn{1}{c}{} &  & \tabularnewline
\end{tabular}

}
\label{TB1}
\end{table*}

For performance evaluation, we employ two commonly used criteria, namely Spearman's rank-order correlation coefficient (SROCC) and Pearson's linear correlation coefficient (PLCC). 
Both SROCC and PLCC range from 0 to 1, and a higher value indicates a better performance.
Following Video Quality Expert Group (VQEG) \cite{antkowiak2000final}, for PLCC, logistic regression is first applied to remove nonlinear rating caused by human visual observation.

\subsection{Implementation Details}
We implemented our model by PyTorch and conducted training and testing on an NVIDIA RTX 2080 GPU. Following the standard training strategy from existing IQA algorithms, we randomly select multiple sample patches from each image and horizontally and vertically augment them randomly. 
Particularly, we select 50 patches randomly with the size of $224\times 224$ pixels from each training image. 
Training patches inherited quality scores from the source image, and we minimize $\mathcal{L}_{total}$ loss over the training set.
We used Adam \cite{kingma2014adam} optimizer with weight decay $5 \times 10^{-4}$ to train our model for at most 5 epochs, with mini-batch size of 53. 
The learning rate is first set to $2 \times 10^{-5}$  and reduced by $10$ after every  epoch.
During the testing stage, 50 patches with $224\times 224$ pixels from the test image are randomly sampled, and their corresponding prediction scores are average pooled to get the final quality score.
We use ResNet50 \cite{he2016deep} for our CNN backbone  unless mentioned otherwise, while it is initialized with Imagenet weights.
We use $N=2$ for number of encoder layers in the Transformer, $d=64$, and set the number of heads $h=16$.
The hyper parameters $\lambda_{1},\lambda_{2},\lambda_{3}$ are empirically set to  $0.5,0.05,1$, respectively.

Following the common practice in NR-IQA,   all experiments use the same setting,  where we first select $10$ different seeds, and  then  use  them  to  split  the  datasets  randomly  to train/test ($80\%/20\%$),  so we have $10$ different splits.   Testing data is not being used during the training.
In the case of synthetically distorted datasets, the split is implemented according to reference images to avoid content overlapping. 
For all of the reported results we
run the experiment $10$ times with different  initialization and report the median SROCC and PLCC values.

% \vspace{-5 pt}
\subsection{Performance Evaluation}
Table \ref{TB1} shows the overall performance comparison in terms of PLCC and SROCC on seven standard image quality datasets, which cover both synthetically and authentically distorted images. 
Furthermore, our model outperforms the existing methods by a significant margin on both LIVEFB and KADID  datasets that are currently the largest datasets for \textit{in-the-wild} images and synthetically distorted images, respectively.
Our model also achieves  competitive results on the smaller datasets.
In the last column, we provide the weighted average  performance across all datasets, using the dataset sizes as  weights for the performances, and we observe that our proposed method outperforms existing methods on both PLCC and SROCC.

In Table \ref{TB2}, we conduct cross dataset evaluations and compare our model to the   competing approaches. 
Training is performed
on one specific dataset, and testing is performed on a different dataset without any finetuning or  parameter adaptation. 
For synthetic image datasets (LIVE, CSIQ, TID2013), we select four   distortion types  (\textit{i.e.,} JPEG, JPEG2K, WN,
and Blur) which all the datasets have in common.
 As shown in Table \ref{TB2}, our proposed method outperforms other algorithms on four datasets among six, which  indicate
the strong generalization power of our approach.

\begin{table}[h]
\centering
\caption{SROCC evaluations on cross  datasets, where bold entries indicate the best performers.}
\resizebox{3.2 in}{!} {
\begin{tabular}{c|cc|c|c|c|c}
\hline 
Train on & \multicolumn{2}{c|}{LIVEFB} & CLIVE & KonIQ & \multicolumn{2}{c}{LIVE}\tabularnewline
\hline 
Test on & KonIQ & CLIVE & KonIQ & CLIVE & CSIQ & TID2013\tabularnewline
\hline 
WaDIQaM\cite{bosse2017deep} & 0.708 & 0.699 & 0.711 & 0.682 & 0.704 & 0.462\tabularnewline
DBCNN\cite{zhang2018blind} & 0.716 & 0.724 & 0.754 & 0.755 & 0.758 & 0.524\tabularnewline
P2P-BM\cite{ying2019patches} & 0.755 & 0.738 & 0.740 & 0.770 & 0.712 & 0.488\tabularnewline
HyperIQA\cite{su2020blindly} & \textbf{0.758} & 0.735 & \textbf{0.772} & 0.785 & 0.744 & 0.551\tabularnewline
TReS (Proposed) & 0.713 & \textbf{0.740} & 0.733 & \textbf{0.786} & \textbf{0.761} & \textbf{0.562}\tabularnewline
\hline 
\end{tabular}
}
\label{TB2}
\end{table}

\begin{figure}[t]
\centering
  \includegraphics [scale=.32]{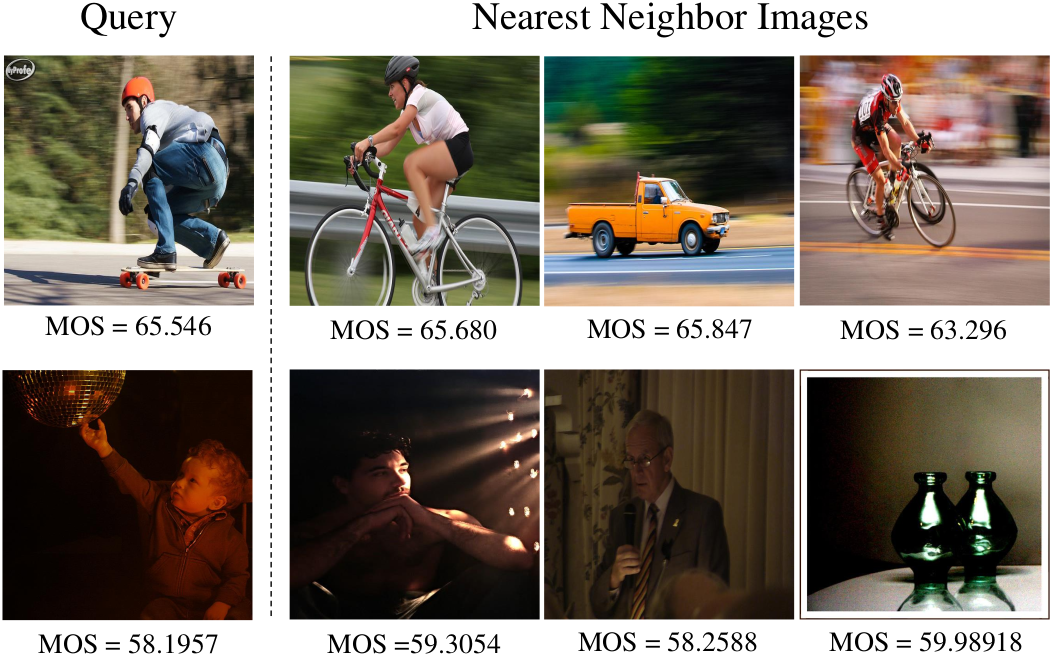}
	\caption{Nearest neighbor retrieval results. In each row, the leftmost
image is a query image, and the rest are the top 3
nearest neighbors, using the latent features learned by our proposed models. The nearest neighbor retrieval process is done on the test portion of the datasets. Images in the first and second rows are taken from   the LIVEFB dataset.}
	\label{F4}
 \end{figure}

We evaluate the latent features learned by our model in Fig. \ref{F4}, where we use the latent features from the last layer of the network for query images and collect the top three nearest neighbor results for the corresponding query. As shown in Fig. \ref{F4}, although we do not explicitly model the content or distortion types in our model, the nearest neighbor samples have similar content or artifacts in terms of perceptual quality and have close subjective scores to each other, which represent the effectiveness of our model in terms of feature representation.
Specifically, in the first row, our model selects images with the same motion blur artifacts as the nearest neighbor samples. In the second row, our model selects images with low lighting condition which follow similar quality conditions as the query image.

\begin{figure}[t]
\centering
  \includegraphics [scale=.33 ]{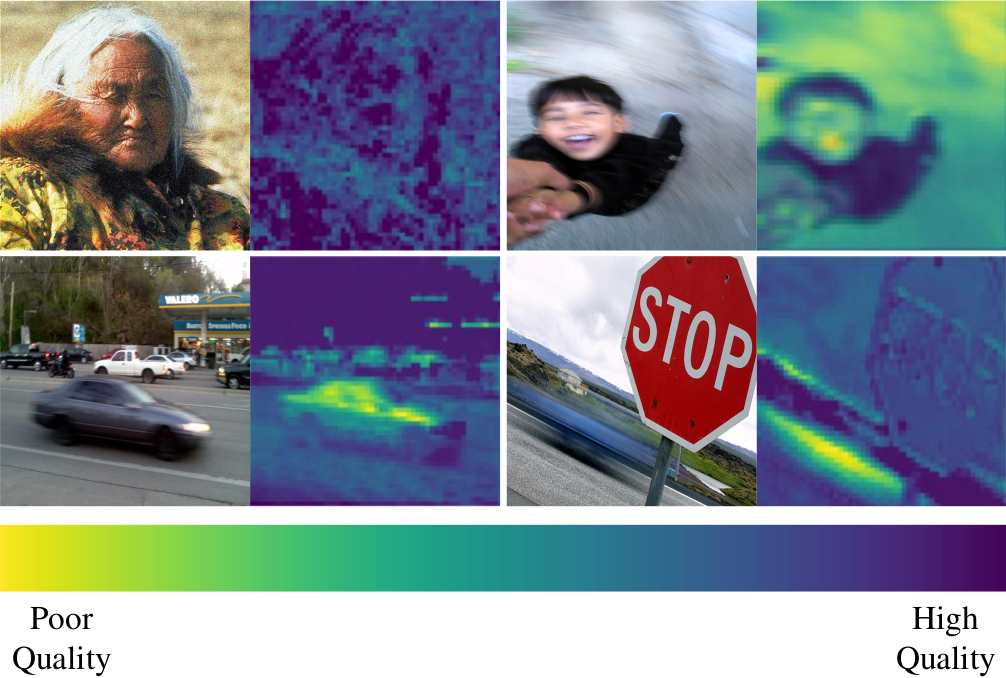}
	\caption{Spatial quality maps generated using the our proposed model. Left: Original Images. Right: Quality maps blended with the
originals using viridis color.}
	\label{F5}
 \end{figure}

Moreover, in Fig. \ref{F5}, we  show the spatial   quality map generated from the layer with the highest activation in our model. The bright regions represent  the poor quality regions in the input images.

\subsection{Ablation Study}
In Table \ref{TB3}, we provide  ablation experiments to illustrate the effect of each  component of our proposed method by comparing the results on KADID and KonIQ datasets.
Furthermore, in Table \ref{TB4}, we evaluate the performance sensitivity of our model for smaller backbones.
For a fair comparison to existing algorithms, we chose Resnet50 for all of the experiments in this paper. 
However, as shown in Table \ref{TB4}, for smaller backbones, our model still achieves comparable results.
As shown in Table \ref{TB4}, for large datasets (\textit{e.g.,} LIVEFB or KonIQ), the performance of our model does not drop significantly and is still competitive when we use a smaller backbone, which demonstrates the learning capacity of our proposed model.

\begin{table}[t]
\centering
\caption{Ablation experiments on the effects of different components for our proposed model.}
\resizebox{3.35 in}{!} {
\begin{tabular}{ccccc|cc|c|c}
\hline 
\multirow{2}{*}{Resnet50} & \multirow{2}{*}{Transformer} & \multicolumn{1}{c}{Psitional} & \multicolumn{1}{c}{Relative} & \multicolumn{1}{c|}{Self} & \multicolumn{2}{c|}{KADID} & \multicolumn{2}{c}{KonIQ}\tabularnewline
\cline{6-9} \cline{7-9} \cline{8-9} \cline{9-9} 
 &  & Encoding & Ranking & Consistency & PLCC & SROCC & \multicolumn{1}{c}{PLCC} & SROCC\tabularnewline
\hline 
\checkmark  &  &  &  &  & 0.809 & 0.802 & 0.873 & 0.851\tabularnewline
\hline 
\checkmark  &  &  & \checkmark  & \checkmark  & 0.822 & 0.820 & 0.896 & 0.884\tabularnewline
\hline 
\checkmark  & \checkmark  &  &  &  & 0.833 & 0.820 & 0.886 & 0.872\tabularnewline
\hline 
\checkmark  & \checkmark  & \checkmark  &  &  & 0.840 & 0.832 & 0.902 & 0.0.895\tabularnewline
\hline 
\checkmark  & \checkmark  & \checkmark  & \checkmark  &  & 0.851 & 0.850 & 0.918 & 0.911\tabularnewline
\hline 
\checkmark  & \checkmark  & \checkmark  & \checkmark  & \checkmark  & 0.858 & 0.859 & 0.928 & 0.915\tabularnewline
\hline 
\end{tabular}
}
\label{TB3}
\end{table}

\begin{table}[t!]
\centering
\caption{Ablation experiments on the performance of our proposed model via different backbones.} 

\resizebox{3.37 in}{!} {
\begin{tabular}{cc|cc|cc|cc}
\hline 
Dataset & Backbone & PLCC & SROCC & Dataset & Backbone & PLCC & SROCC\tabularnewline
\hline 
\multirow{3}{*}{CLIVE} & Resnet-50 & 0.877 & 0.846 & \multirow{3}{*}{CSIQ} & Resnet-50 & 0.942 & 0.922\tabularnewline
 & Resnet-34 & 0.855 & 0.830 &  & Resnet-34 & 0.924 & 0.920\tabularnewline
 & Resnet-18 & 0.859 & 0.822 &  & Resnet-18 & 0.911 & 0.914\tabularnewline
\hline 
\multirow{3}{*}{KonIQ} & Resnet-50 & 0.928 & 0.915 & \multirow{3}{*}{TID2013} & Resnet-50 & 0.883 & 0.863\tabularnewline
 & Resnet-34 & 0.922 & 0.909 &  & Resnet-34 & 0.847 & 0.813\tabularnewline
 & Resnet-18 & 0.909 & 0.898 &  & Resnet-18 & 0.843 & 0.810\tabularnewline
\hline 
\multirow{3}{*}{LIVEFB} & Resnet-50 & 0.625 & 0.560 & \multirow{3}{*}{KADID} & Resnet-50 & 0.858 & 0.859\tabularnewline
 & Resnet-34 & 0.619 & 0.554 &  & Resnet-34 & 0.851 & 0.855\tabularnewline
 & Resnet-18 & 0.611 & 0.550 &  & Resnet-18 & 0.840 & 0.848\tabularnewline
\hline 
\end{tabular}
}
\label{TB4}
\end{table}

\subsection{Failure Cases and Discussion }
\begin{figure}[h]
\centering
  \includegraphics [scale=.41 ]{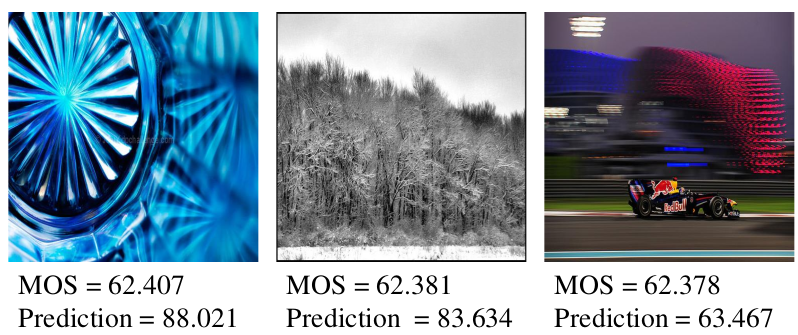}
	\caption{Failure cases, where the predictions
are different from the subjective scores (MOS).}
	\label{F6}
 \end{figure}

In Fig. \ref{F6}, we show examples where our method fails to predict the image quality in agreement with the human subjective scores.
All images in Fig. \ref{F6} have close ground truth scores, and our model predicted different scores for each image.
From the modeling aspect, we think one  reason for such a failure is that IQA models have to address the IQA task as either a regression and/or classification problem (simply because the existing datasets provide only the quality score(s) for each image). Recently, LIVEFB \cite{ying2019patches} provides  patch-wise quality scores for local patches of each image and shows that incorporating patch scores leads to better performance. 
As a future direction, we think what is missing from the existing IQA datasets is a description of the reasoning process from the subjects to explain the reason behind their selected quality score; this can help the future models   be able to model the HVS and reasoning behind the assigned quality scores in a better way for a more precise perceptual quality assessment.
On the other hand, from the subjective scores perspective, the subjects may be less forgiving of the blur artifact and grayscale images, so those artifacts have drawn their attentions similarly.  However, our model differentiates between different perceptual cues (color, sharpness, blurriness), which can explain the differences in the scores.

% \vspace{-0.6 cm}
\section{Conclusion}
In this work, we present an NR-IQA algorithm that works based on a hybrid combination of CNNs and Transformers features to utilize both local and non-local feature representation of the input image.
We further propose a  relative ranking loss that takes into account the relative ranking information among the images.
Finally, we   exploit an additional self-consistency loss to improve the robustness of our proposed method. 
Our experiments show that our proposed method performs well on several IQA datasets covering  synthetically and authentically distorted images.

\clearpage
{\small
\bibliographystyle{ieeetr}
\bibliography{egbib}
}

\clearpage

\begin{figure*}[h]
\centering
  \includegraphics [scale=.32 ]{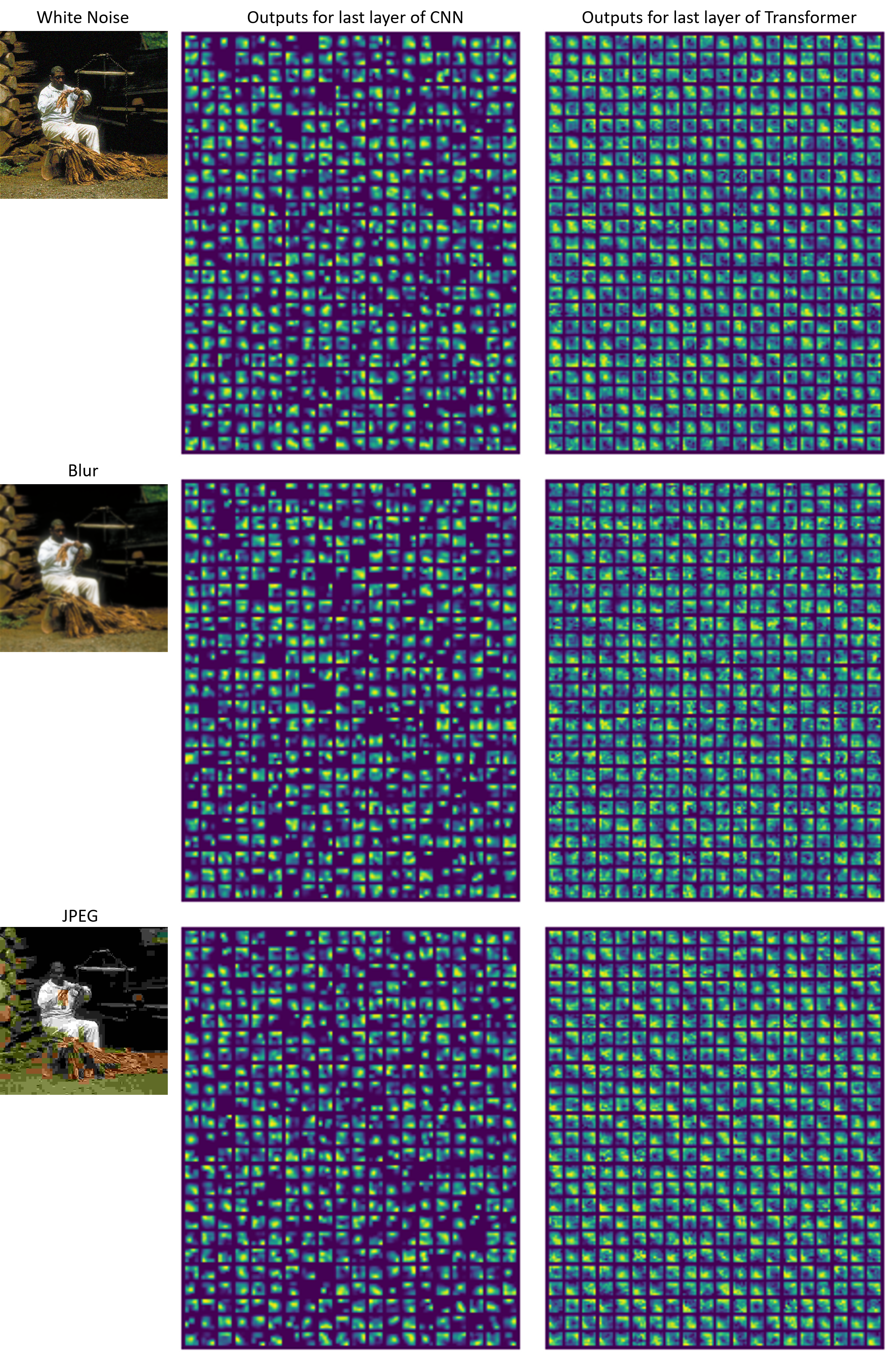}
	\caption{Visualization of features from the last layer of CNN and Transformer.
	Our CNN and Transformer models capture different information for each image as well as across different distortion types.
	}
	\label{F9}
 \end{figure*}

Fig. \ref{F9} shows the scatter plots of our model's predictions \textit{v.s.}  subjective ratings (MOS/DMOS) for all seven datasets. As shown in Fig. \ref{F9}, the recently proposed FBLIVE dataset is the most challenging one. Although we achieve state-of-the-art on FBLIVE dataset comparing to existing algorithms, there is still a long way to go for understanding how the human vision system evaluates the quality of images in-the-wild.

A fair question is why not using other types of transformation instead of horizontal flipping and how they will affect the performance if we use them for self-consistency.
In Table \ref{TB3s}, we provide an ablation study where we consider other types of transformations to enforce self-consistency.

\begin{table}[h]
\centering
\caption{SROCC results for ablation study for different equivariant transformations for the self-consistency loss.}
\resizebox{2.45 in}{!} {
\begin{tabular}{c|c|c}
\hline 
Components & KADID & KonIQ\tabularnewline
\hline 
Baseline & 0.850 & 0.911\tabularnewline
Horizontal Flipping & 0.859 & 0.915\tabularnewline
Vertical Flipping & 0.852 & 0.882\tabularnewline
Rotation 90 Degree & 0.851 & 0.891\tabularnewline
Random Translation 16-20 pixels & 0.854 & 0.913\tabularnewline
Random Crop & 0.832 & 0.872\tabularnewline
Horizontal Flipping + Translation & 0.860 & 0.916\tabularnewline
\hline 
\end{tabular}

}
\label{TB3s}
\end{table}

Based on our experiments, for synthetically distorted datasets horizontal and vertical flipping, rotation, and translation can also improve the performance.
For authentically distorted datasets we observe that only horizontal flipping and translation yield performance improvement. 
These experiments can explain that for synthetic datasets the artifacts play an important role and different transformations can help the model captures the artifacts better and become independent of the content information. However, for authentic quality assessment, the quality score is a combination of different factors and therefore some of the transformations can hurt the performance instead of helping. 
For example, to a viewer, a rotated version of the image will not have the same authentic quality score as the original one, therefore, enforcing the self-consistency would not be beneficial.

Also, we would like to emphasize that although our self-consistency will not add any computation to the interface time, it will require more GPU memory during the training, therefore it will be computationally expensive to apply multiple transformations at the same time. Nonetheless, in the last row of Table \ref{TB3s}, we show that a combination of horizontal flipping and translation can further improve the results.
As future work, we would like to consider improving the computational complexity of our self-consistency idea.

Last but not least, we claim in the paper that the features generated from CNNs and Transformers represent two different aspects of an image. In Fig. \ref{F9}, we visualize the features of the last layer of our CNN model (second column) and our Transformer model (third column) for images with the same content but different artifacts. We can observe that for each image the CNN features and Transformer features represent completely different information. Moreover, we can observe that the features from the CNN  (or Transformer) layer for different artifacts are also different, which proves the power of our model in capturing different informative information for each image.

%approach
%why just hrorizental filipping, we also agree that other transformations should also take into account, however, it is not clear how the human subjective scores will change for other transformations. Nonethelss, we performe an ablation study to show...

% \section*{More Details on the Relative Ranking Loss}

\section*{FQAs}

\textbf{1)} What is the difference between self-consistency and ensembling? and will the self-consistency increase the interface time?
In ensampling methods, we need to have several models (with different initializations) and ensemble the results during the training and testing, but in our self-consistency model, we enforce one model to have consistent performance for one network during the training while the network has an input with different transformations.
Our self-consistency model has the same interface time/parameters in the testing similar to the model without self-consistency. In other words, we are not adding any new parameters to the network and it won't affect the interface.

\textbf{2}) What is the difference between self-consistency and augmentation?
In augmentation, we augment an input and send it to one network, so although the network will become robust to different augmentation, it will never have the chance of enforcing the outputs to be the same for different versions of an input \textit{at the same time}. In our self-consistency approach, we force the network to have a similar output for an image with a different transformation (in our case horizontal flipping) which leads to more robust performance. Please also note that we still use augmentation during the training, so our model is benefiting from the advantages of both augmentation and self-consistency. Also, please see Fig. 1 in the main paper, where we showed that models that used augmentation alone are sensitive to simple transformations.

\textbf{3}) Why does the relative ranking loss apply to the samples with the highest and lowest quality scores, why not applying it to all the samples?
1) We did not see a significant improvement by applying our ranking loss to all the samples within each batch compared to the case that we just use extreme cases. 
2) Considering more samples  lead to more gradient back-propagation and therefore more computation during the training which causes slower training.

\end{document}